\documentclass[12pt]{article}
\usepackage{graphicx}
\usepackage{amsfonts}
\usepackage{amsmath}
\usepackage{mathrsfs}
\def\beq{\begin{equation}}
\def\eeq{\end{equation}}
\def\bea{\begin{eqnarray}}
\def\eea{\end{eqnarray}}
\def\nn{\nonumber}
\def\noi{\noindent}
\def\ba{\begin{array}}
\def\ea{\end{array}}

\def\v{\vert}
\def\l{\langle}
\def\r{\rangle}
\def\mc{\multicolumn}
\def\one{1\hskip -1mm{\rm l}}
\DeclareMathOperator{\wt}{wt}

\begin{document}

\begin{center}
{\large \bf \sf
Boson-fermion duality in $SU(m|n)$ supersymmetric \\ 
Haldane-Shastry spin chain}

\vspace{1.3cm}

{\sf B. Basu-Mallick$^1$\footnote{e-mail: bireswar.basumallick@saha.ac.in}, 
Nilanjan Bondyopadhaya$^1$\footnote{e-mail: 
nilanjan.bondyopadhaya@saha.ac.in},
Kazuhiro Hikami $^2$\footnote{e-mail: hikami@phys.s.u-tokyo.ac.jp}
and Diptiman Sen$^3$\footnote{e-mail: diptiman@cts.iisc.ernet.in}}
\bigskip

{\em $^1$Theory Group, Saha Institute of Nuclear Physics, \\
1/AF Bidhan Nagar, Kolkata 700 064, India} 
\bigskip

{\em $^2$Department of Physics, Graduate School of Science, University 
of Tokyo, \\
Hongo 7-3-1, Bunkyo, Tokyo 113-0033, Japan}
\bigskip

{\em $^3$Centre for High Energy Physics, Indian Institute of Science, \\
Bangalore 560 012, India}
\end{center}

\vskip 2 cm 
\medskip
\noi {\bf Abstract} 
\medskip

By using the $Y(gl(m|n))$ super Yangian symmetry of the $SU(m|n)$ 
supersymmetric Haldane-Shastry spin chain, we show that the partition function
of this model satisfies a duality relation under the exchange of bosonic and
fermionic spin degrees of freedom. As a byproduct of this study of the duality
relation, we find a novel combinatorial formula for the super Schur polynomials
associated with some irreducible representations of the $Y(gl(m|n))$ Yangian
algebra. Finally, we reveal an intimate connection between the global $SU(m|n)$
symmetry of a spin chain and the boson-fermion duality relation.
\bigskip

\newpage 

\baselineskip=18pt
\noi \section{Introduction}
\renewcommand{\theequation}{1.{\arabic{equation}}}
\setcounter{equation}{0}

\medskip
The appearance of boson-fermion duality in lower 
dimensional quantum field theoretical models and many-particle systems
has attracted a lot of attention in recent years [1-8]. 
The equivalence of bosonic sine-Gordon model with fermionic massive
Thirring model is a classic example of boson-fermion duality in 
the context of one-dimensional field theoretical models [1].
Such a duality is a consequence of the fact that spin and statistics become 
essentially irrelevant notions in one spatial dimension, and thus the bosonic 
and fermionic theories can be related to each other through a duality 
transformation. In the context of many-particle systems in one dimension, it 
has been found that a bosonic and a fermionic model, with distinct point 
interactions and related by coupling constant inversion, share the same 
spectrum [2,3]. The signature of boson-fermion duality has also been observed 
in the setting of quantum many-body systems like the Tomonaga-Luttinger liquid
theory of one-dimensional systems of interacting fermions, where the 
low-lying excitations are describable through bosonic degrees of freedom 
[4-6]. Recently, an exact bosonization method has been applied to study this 
problem in the non-interacting case even beyond the regime of validity of 
the low-energy approximation [7]. 

Duality relation has also been explored in the 
context of supersymmetric quantum integrable spin models,
where bosonic and fermionic spin degrees of freedom appear simultaneously. 
It may be noted that, such exactly solvable one dimensional quantum 
spin chains have a close relation with correlated systems in 
condensed matter physics, where holes moving in the dynamical background of 
spins behave as bosons and spin$-\frac{1}{2}$ electrons behave as fermions 
[9,10]. Recent studies also reveal some interesting connection of these 
supersymmetric spin chains with loop models [11]. The Haldane-Shastry (HS) 
spin chain and the Polychronakos spin chain are two well known examples of 
quantum integrable models with long range interaction, for which the exact 
spectra can be computed analytically even for finite number of lattice sites 
[12-16]. Supersymmetric extensions of these spin chains and related exactly 
solvable models have also been studied intensively [17-24]. 
In particular, by using the freezing trick [13,19], the exact partition 
function of the $SU(m|n)$ supersymmetric Polychronakos spin chain 
has been derived in a simple 
form [21,22]. Furthermore, it has been shown analytically that, the partition 
function of this supersymmetric Polychronakos spin chain satisfies a duality 
transformation under the exchange of bosonic and fermionic spin degrees of 
freedom. So it is natural to enquire whether a similar boson-fermion duality 
relation exists in the case of the $SU(m|n)$ supersymmetric HS spin chain.

The Hamiltonian of the $SU(m|n)$ HS model, with $N$ number of lattice sites 
uniformly distributed on a circle, is given by
\beq H^{(m|n)}=\frac{1}{2}\sum_{1\leq j <k \leq N}
\frac{(1+\hat{P}_{jk}^{(m|n)})}{\sin^2(\xi_j - \xi_k )} \, , \label{a1} \eeq
where $\xi_j=j\pi /N$, and $\hat{P}_{jk}^{(m|n)}$ is the supersymmetric
exchange operator (its definition is given in Section 2) which interchanges
the `spins' on the $j$-th and $k$-th lattice sites. In analogy with the 
nonsupersymmetric case [25], the exact partition function of the $SU(m|n)$ HS 
spin chain (\ref{a1}) has been computed recently by applying the freezing 
trick [23]. It has also been conjectured that this partition function, which 
is denoted by $Z_N^{(m|n)}(q)$, satisfies a duality relation of the form 
\beq Z_N^{(m|n)}(q)=q^{\frac{N(N^2-1)}{6}}Z_N^{(n|m)}(q^{-1}) \, ,
\label{a2} \eeq
where $q \equiv e^{-\frac{1}{k_BT}}$. With the help of a symbolic software 
package like Mathematica, one can easily check the validity of this 
conjecture for a wide range of values of $m$, $n$ and $N$. However, an 
analytical proof of this conjecture for all possible values of $m$, $n$ and 
$N$ has been lacking till now. The boson-fermion duality for the
supersymmetric HS spin chain can also be studied at the level of the
corresponding spectrum. Comparing the coefficients of the same powers of $q$ 
on the two sides of \mbox{Eqn. (\ref{a2})}, one finds that
the spectrum of $H^{(m|n)}$ (\ref{a1}) is related to that of 
$H^{(n|m)}$ through an inversion and an overall shift of all 
energy levels. Such a relation between the spectra of the $SU(m|n)$ 
and $SU(n|m)$ HS spin chains was first empirically observed by Haldane 
on the basis of results obtained by numerical diagonalization [17]. 

In this article, we aim to provide an analytical proof for 
the duality relation (\ref{a2}). To this end, it may be noted that
the $SU(m|n)$ supersymmetric spin Calogero-Sutherland (CS) 
model has a $Y(gl(m|n))$ super Yangian symmetry [26]. 
Since the $SU(m|n)$ HS spin chain can be obtained 
by taking the freezing limit of the $SU(m|n)$ spin CS model, the Hamiltonian 
(\ref{a1}) also exhibits the $Y(gl(m|n))$ super Yangian symmetry [17]. 
It is well known that, a family of irreducible representations of this 
$Y(gl(m|n))$ quantum group can be labeled by some super skew Young diagrams 
and the corresponding Schur polynomials. Interestingly, such super Schur 
polynomials obey a duality relation under the exchange of bosonic and 
fermionic variables [27,28]. This duality relation will play a key role in our 
approach for proving the boson-fermion duality relation in the case of the 
$SU(m|n)$ HS spin chain. In Sec. 2 of this article, we briefly review the 
super Yangian symmetry for the $SU(m|n)$ HS spin chain and also give a 
simple alternative proof of the duality relation satisfied by the super 
Schur polynomials. In Sec. 3, we find a novel combinatorial formula for 
these super Schur polynomials, which allows us to establish a connection 
between these polynomials and the partition function of the $SU(m|n)$ HS 
spin chain. By exploiting the above mentioned connection, we give an 
analytical proof of the boson-fermion duality relation (\ref{a2}) in Sec. 4. 
In Sec. 5, we explore the possibility of constructing a class of quantum 
integrable as well as nonintegrable spin chains which would satisfy the 
boson-fermion duality relation. Sec. 6 is the concluding section.

\noi \section{$Y(gl(m|n))$ super Yangian symmetry of $SU(m|n)$ HS spin chain}
\renewcommand{\theequation}{2.{\arabic{equation}}}
\setcounter{equation}{0}

For the purpose of defining the super exchange operator in the Hamiltonian 
(\ref{a1}) of the $SU(m|n)$ HS spin chain, let us consider a set of operators 
like $C_{j \alpha}^\dagger$($C_{j \alpha}$) which creates (annihilates)
a particle of species $\alpha$ on the $j$-th lattice site. These creation 
(annihilation) operators are assumed to be bosonic when $\alpha \in 
\{1,2,....,m\}$ and fermionic when $\alpha \in \{m+1,m+2,....,m+n\}$.
Thus, the parity of $C_{j \alpha}^\dagger$($C_{j \alpha}$) is defined as 
\bea && p(C_{j \alpha})=p(C_{j \alpha}^\dagger)=0 ~
\mathnormal{for}~ \alpha \in \{1,2,....,m \} \, , \nn \\
&& p(C_{j \alpha})=p(C_{j \alpha}^\dagger)=1 ~
\mathnormal{for}~ \alpha \in \{ m+1,m+2,....,m+n \} \, . \nn \eea
These operators satisfy (anti-) commutation relations like 
\beq [C_{j \alpha},C_{k \beta}]_{\pm}=0 \, ,~ 
[C_{j \alpha}^\dagger,C_{k \beta}^\dagger]_{\pm}=0 \, , ~
[C_{j \alpha},C_{k \beta}^\dagger]_{\pm}=\delta_{jk}\delta_{\alpha \beta} \, ,
\label{b1} \eeq
where $[A,B]_{\pm} \equiv AB- (-1)^{p(A)p(B)}BA$.
Next, we focus our attention on a subspace of the related Fock space, for
which the total number of particles per site is always one:
\beq \sum_{\alpha=1}^{m+n} C_{j\alpha}^{\dagger} C_{j\alpha}=1, \label{b2} \eeq
for all $j$. On the above mentioned subspace, one can define the 
supersymmetric exchange operators as 
\beq \hat{P}_{jk}^{(m|n)} \equiv \sum_{\alpha,\beta=1}^{m+n} 
C_{j \alpha}^\dagger C_{k \beta}^\dagger C_{j \beta}C_{k \alpha} \, ,
\label{b3} \eeq
where $1 \leq j <k \leq N$. Inserting these exchange operators in (\ref{a1}),
one obtains the Hamiltonian of the $SU(m|n)$ HS spin chain.

Next, we shall briefly review the super Yangian symmetry of this $SU(m|n)$ 
HS spin chain. Let $\mathbf{V}$ be an $(m+n)$-dimensional auxiliary graded 
vector space, through which the graded Yang--Baxter equation will be defined.
We set $\mathbf{B} = \mathbf{B}_+ \sqcup \mathbf{B}_-$, where
\beq \mathbf{B}_+= \{ \epsilon_1, \dots, \epsilon_m \} \, , ~~~~~~
\mathbf{B}_- = \{ \epsilon_{m+1}, \dots, \epsilon_{m+n} \} . \label{b4} \eeq
The generators $\mathbf{E}^{\alpha\beta}$ of the $gl(m|n)$ Lie algebra 
satisfy (anti-)commutation relations given by 
\begin{equation*}
[\mathbf{E}^{\alpha\beta} , \mathbf{E}^{\gamma\delta}]_\pm = 
\delta^{\beta \gamma} \mathbf{E}^{\alpha \delta} - (-1)^{ \left( p(\alpha) + 
p(\beta) \right) \left( p(\gamma) + p(\delta) \right)}
\delta^{\alpha \delta} \mathbf{E}^{\gamma \beta} \, , \end{equation*}
where $p(\alpha)=0$ (resp. $p(\alpha)=1$) if $\epsilon_\alpha \in 
\mathbf{B}_+$ (resp. $\epsilon_\alpha \in \mathbf{B}_-$).
With these generators, we define the graded permutation operator as
\beq \mathbf{P} = \sum_{\alpha, \beta=1}^{m+n} (-1)^{p(\beta)}
\mathbf{E}^{\alpha \beta} \otimes \mathbf{E}^{\beta \alpha} \, . \eeq
Let $\{ e^\alpha \}$ be a set of basis vectors of the auxiliary vector space 
$\mathbf{V}$. The generators $\mathbf{E}^{\alpha\beta}$ and the graded 
permutation operator $\mathbf{P}$ act on such basis vectors and their direct 
products as 
\begin{gather*}
\mathbf{E}^{\alpha \beta} e^\gamma = \delta^{\beta \gamma} e^\alpha \, , \\
\mathbf{P} \, e^\alpha \otimes e^\beta = (-1)^{p(\alpha) p(\beta)}
e^\beta \otimes e^\alpha \, . \end{gather*}
We have the rational solution of the graded Yang-Baxter equation given by 
\beq R(u) = u - \hbar \, \mathbf{P} \, , \label{b6} \eeq
where $u$ is the spectral parameter. The $Y(gl(m|n))$ super Yangian [29] 
is associated to this $R$-matrix. Namely,
\beq R(u-v) \, \overset{1}{\mathbf{T}}(u) \, \overset{2}{\mathbf{T}}(v)
= \overset{2}{\mathbf{T}}(v) \, \overset{1}{\mathbf{T}}(u) \, R(u-v) \, ,
\label{b7} \eeq
where 
$\overset{1}{\mathbf{T}}(u) \equiv {\mathbf{T}}(u) \otimes \one $, 
$\overset{2}{\mathbf{T}}(v) \equiv \one \otimes {\mathbf{T}}(v)$ and 
${\mathbf{T}}(u)$ is defined as 
\begin{equation*}
\mathbf{T}(u) = \sum_{\alpha, \beta=1}^{m+n} (-1)^{p(\alpha)} \,
T^{\alpha \beta}(u) \, \mathbf{E}^{\alpha \beta} \, . \end{equation*}
Computing the tensor products through the rule 
\begin{equation*}
\left( a_1 \otimes b_1 \right) \, \left( a_2 \otimes b_2 \right)
= (-1)^{p(a_2) p(b_1)} \, a_1 \, a_2 \otimes b_1 \, b_2 \, , \end{equation*}
one can express \mbox{Eqn. (\ref{b7})} as
\bea
&& \left [ T^{\alpha \beta}(u) , T^{\gamma \delta}(v) \right]_\pm ~~~~\nn \\
&& ~~~= \frac{\hbar}{u-v} \, (-1)^{p(\alpha) p(\beta) + p(\gamma) p(\beta) 
+ p(\alpha) p(\gamma)} \left( T^{\gamma \beta}(u) \, T^{\alpha \delta}(v) -
T^{\gamma \beta}(v) \, T^{\alpha \delta}(u) \right). \label{b8} \eea
The Yangian currents $T^{\alpha \beta}(u)$ may be expanded in powers of 
the spectral parameter as 
\beq T^{\alpha \beta}(u) = \delta^{\alpha \beta} + \hbar \, \sum_{n=0}^\infty
(-1)^{p(\alpha)} \, \frac{T_n^{\alpha \beta}}{u^{n+1}} \, , \label{b9} \eeq
and the $Y(gl(m|n))$ algebra (\ref{b8}) can also be expressed in a spectral 
parameter independent way through the generators $T_n^{\alpha \beta}$. 

The $Y(gl(m|n))$ super Yangian symmetry has been realized explicitly in the 
case of the $SU(m|n)$ HS spin chain [17,30]. Suitable combinations of the 
generators $T_0^{\alpha \beta}$ and $T_1^{\alpha \beta}$ yield conserved 
quantities of the Hamiltonian (\ref{a1}) in the form 
\bea && ~~~~~~~~~~~Q_0^{\alpha \beta} = \sum_{j=1}^N \left( 
C_{j \alpha}^\dagger C_{j \beta} - \frac{1}{ m + n} \delta_{\alpha \beta} 
\right) , \nn \hskip 5.5cm (2.10a) \\
&& ~~~~~~~~~~~Q_1^{\alpha \beta} = \frac{1}{2} \sum_{j \neq k} 
\sum_{\gamma=1}^{m+n} \cot \left( \frac{j-k}{N} \pi \right)
C_{j\alpha}^\dagger C_{k \gamma}^\dagger C_{j \gamma} C_{k \beta} \, .
\nn \hskip 3.5cm (2.10b) \eea
\addtocounter{equation}{1}
It is well known that the $Y(gl(m|n))$ Yangian algebra is effectively generated
by the lowest two generators $T_0^{\alpha \beta}$ and $T_1^{\alpha \beta}$.
Consequently, by using the commutation relations among conserved quantities 
like $Q_0^{\alpha \beta}$ and $Q_1^{\alpha \beta}$, one can obtain the 
complete $Y(gl(m|n))$ Yangian symmetry of the $SU(m|n)$ HS spin chain. 



Next, we shall prepare the super Schur polynomials (see e.g. [31]) which are 
closely related to a class of irreducible representations of $Y(gl(m|n))$ 
Yangian algebra. For $\mathbf{B}$, we set a usual ordering as
\begin{equation*}
\epsilon_1 \prec \epsilon_2 \prec \cdots \prec \epsilon_{m+n} \, .
\end{equation*}
The Young tableaux $T$ is obtained by filling the numbers $1, 2, \dots, m+n$ 
in a given Young diagram $\lambda$ by the rules:
\begin{itemize}
\item Entries in each row are increasing, allowing the repetition of elements
in $\{i | \epsilon_i \in \mathbf{B}_+ \}$, but not permitting the repetition 
of elements in $\{i | \epsilon_i \in \mathbf{B}_- \}$,
\item Entries in each column are increasing, allowing the repetition of
elements in $\{i | \epsilon_i \in \mathbf{B}_- \}$, but not permitting the 
repetition of elements in $\{i | \epsilon_i \in \mathbf{B}_+ \}$. 
\end{itemize}
The super Schur polynomial corresponding to the Young diagram $\lambda$ is 
then defined as
\beq S_\lambda( x, y)
= \sum_{ \text{tableaux $T$ of shape $\lambda$}} e^{\wt(T)} \, . 
\label{define_super_Schur} \eeq
Here the weight $\wt(T)$ of the Young tableaux $T$ is given by 
\begin{equation*}
\wt(T) = \sum_\alpha m_\alpha \, \epsilon_\alpha \, , \end{equation*}
where $m_\alpha$ denotes the number of $\alpha$ in $T$, and we use 
the notations 
\begin{align*}
x_i & \equiv e^{\epsilon_i} \, \text{for $\epsilon_i \in \mathbf{B}_+$} \, ,
& y_i & \equiv e^{\epsilon_{m+i}} \, 
\text{for $\epsilon_{m+i} \in \mathbf{B}_-$} \, , \end{align*}
along with $x\equiv \{x_1,\dots,x_m\}$, $y\equiv \{y_1,\dots,y_n\}$. 


A skew Young diagram $\lambda /\mu$ is obtained by removing a smaller Young 
diagram $\mu$ from a larger one $\lambda$ that contains it [32]. The super 
Schur polynomial $S_{\lambda / \mu}(x,y)$ corresponding to such 
skew Young diagram $\lambda /\mu$ can also be defined 
combinatorially as in \eqref{define_super_Schur}. 
Let $\lambda^\prime$ denote the conjugate of the Young diagram $\lambda$ 
(the conjugate of a Young diagram is obtained by flipping it over its main 
diagonal). It is evident that the rows of a conjugate diagram are mapped to 
the columns of the original diagram and vice versa. It is worth noting that 
the rule for filling up a \emph{row} of a super Young tableaux, as stated 
before \mbox{Eqn. \eqref{define_super_Schur}}, is transformed to the rule 
for filling up a \emph{column} of a super Young tableaux (and vice versa)
provided we substitute the elements 
$\{i | \epsilon_i \in \mathbf{B}_+ \}$ in place of elements 
$\{i | \epsilon_i \in \mathbf{B}_- \}$ (and vice versa).
Due to such a \emph{duality} of the rules for filling numbers in the
case of a skew Young tableaux, we easily obtain 
\beq S_{\lambda / \mu}(x, y ) = S_{\lambda^\prime / \mu^\prime} (y,x) \, .
\label{b12} \eeq
This duality relation between two super Schur polynomials associated with
a skew Young diagram and its conjugate diagram 
has also been found earlier through a different approach [27,28]. 

Hereafter we shall consider only connected super skew
Young diagrams which do not contain any $2\times 2$ square box. Such a 
skew Young diagram is also called a `border strip' and may be denoted by
$\langle m_1, m_2, \dots, m_r \rangle$: 
\vskip .02cm 
\begin{figure}[h]
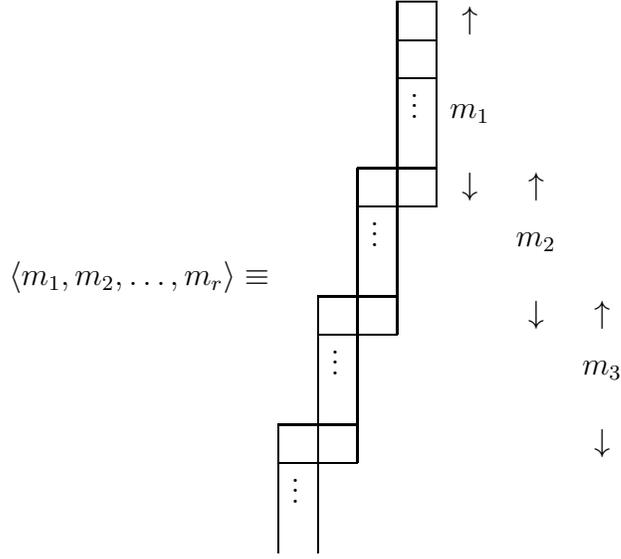

\beq \langle m_1, m_2, \dots , m_r \rangle \equiv 
\begin{array}{*{4}{|p{\arraycolsep}}|ccc}
\cline{4-4} \multicolumn{3}{c|}{} & & \uparrow & & \\
\cline{4-4} \multicolumn{3}{c|}{} & & & & \\
\cline{4-4} \multicolumn{3}{c|}{} & \vdots & m_1 & & \\
\cline{3-4} \multicolumn{2}{c|}{} & & & \downarrow & \uparrow & \\
\cline{3-4} \multicolumn{2}{c|}{} & \vdots & \multicolumn{1}{|c}{} & & m_2 & \\
\cline{2-3} \multicolumn{1}{c|}{} & & &\multicolumn{1}{|c}{} & & \downarrow
& \uparrow \\
\cline{2-3} \multicolumn{1}{c|}{} & \vdots & \multicolumn{2}{|c}{} & & & m_3 \\
\cline{1-2} & & \multicolumn{2}{|c}{} & & & \downarrow \\
\cline{1-2} \vdots & \multicolumn{6}{|c}{} \\
\end{array} \nn \eeq
\caption{Shape of the border strip $\langle m_1, m_2, \dots , m_r \rangle$}
\end{figure}
\noi These border strips will play a key role for our purpose due to their 
connection with `motifs' [30,32,33], which represent irreducible 
representations of Yangian algebra and span the Fock space of Yangian 
invariant spin systems. The motif $\delta$ for an $N$-site super spin chain 
is given by an $N-1$ sequence of 0's and 1's, $\delta = (\delta_1, \delta_2, 
\dots , \delta_{N-1})$ with $\delta_j \in \{ 0, 1\}$. There exists a 
one-to-one map from a motif $\delta$ to the border strip $\langle m_1,m_2, 
\dots, m_r \rangle$; we read a motif $\delta =(\delta_1, \delta_2, \dots )$ 
from the left, and add a box under (resp. left) the box when we encounter 
$\delta_j=1$ (resp. $\delta_j=0$). For example, the motif $(10110)$ leads 
to the border strip $\l 2,3,1 \r$. The inverse mapping from a border strip 
to a motif can also be defined in a straightforward way. 

Finally we discuss a convenient way of expressing the super Schur polynomials 
associated with border strips through the supersymmetric elementary functions.
Let us assume that the polynomial $e_\ell(x)$ represents the sum of all 
monomials $x_{i_1}x_{i_2}\cdots x_{i_\ell}$ for all strictly increasing 
sequences $1 \leq i_1 <i_2< \dots < i_\ell \leq m$, while the polynomial
$h_\ell(y)$ represents the sum of all distinct monomials of degree
$\ell$ in the variables $y$. We have the following 
generating functions for polynomials $e_\ell(x)$ and $h_\ell(y)$:
\begin{equation*} 
~~~~~~~~~~~ \prod_{i=1}^m ( 1 + t \, x_i) = \sum_{\ell=0}^m e_\ell(x) \, 
t^\ell \, , ~~~~\prod_{i=1}^n \frac{1}{ 1 - t \, y_i } =
\sum_{\ell=0}^\infty h_\ell(y) \, t^\ell \, . ~~~~~~~~~~~~~~(2.13a,b)
\end{equation*} 
\addtocounter{equation}{1}
The supersymmetric elementary function $E_{j}(x,y)$ 
($\, \equiv S_{[1^{j}]}(x,y)$) may be written through the polynomials 
$e_\ell(x)$ and $h_\ell(y)$ as [22,34]
\beq E_{j}(x,y) \, = \, \sum_{\ell=0}^{j} e_\ell(x) \, h_{j-\ell}(y) \, .
\label{b14} \eeq
By using these supersymmetric elementary functions, one can express the 
super Schur polynomial corresponding to the border strip $\langle m_1 , 
m_2 , \dots, m_r \rangle$ in the form of a determinant given by [22,32]
\beq S_{\langle m_1 , m_2, \dots, m_r \rangle}(x,y) =
\begin{vmatrix}
E_{m_r} & E_{m_r + m_{r-1}} & \dots & \dots & E_{m_r + \dots + m_1} \\
1 & E_{m_{r-1}} & E_{m_{r-1} + m_{r-2}}& \dots & E_{m_{r-1}+ \dots + m_1} \\
0 & 1 & E_{m_{r-2}}& \dots & \vdots \\
\vdots & \ddots & \ddots & \ddots& \vdots \\
0 & \dots & 0 & 1 & E_{m_1}
\end{vmatrix} . \label{b15} \eeq
Expansion of the above determinant along its first row yields a 
recursion relation for the super Schur polynomials as
\beq S_{\l m_1, m_2, \dots,m_r \r }(x,y)
=\sum_{s=1}^{r} (-1)^{s+1} E_{m_r+m_{r-1}+ \dots+m_{r-s+1}}(x,y)
\cdot S_{\l m_1, m_2, \dots,m_{r-s}\r }(x,y) \, , \label{b16} \eeq
where $S_{\l 0 \r }(x,y)=1$. This recursion relation will play 
an important role in our analysis in the next section. 

\noi \section{Partition function of $SU(m|n)$ HS spin chain and super Schur 
polynomials}
\renewcommand{\theequation}{3.{\arabic{equation}}}
\setcounter{equation}{0}

Here our aim is to make a connection between the partition function of the
$SU(m|n)$ supersymmetric HS spin chain and the super Schur polynomials 
associated with the border strips. The partition function for the Hamiltonian 
(\ref{a1}) of the $SU(m|n)$ HS spin chain is found to be [23]
\beq Z_{N}^{(m|n)}(q)= \sum_{r=1}^N \sum_{\{m_1, m_2, \dots,m_r \} \in 
\mathcal {P}_N(r)} \Bigl( \prod_{i=1}^r d^{(m|n)}_{m_i} \Bigr) \, 
F_{m_1, m_2, \dots,m_r}(q) \, , \label{c1} \eeq
where $q \equiv e^{-1/k_BT}$, $~d^{(m|n)}_{m_i}$ is given by
\beq d^{(m|n)}_{m_i}= \sum_{k=0}^{{\rm min}(m_i,\, m)}
C_k^m \, C_{m_i-k}^{m_i-k+n-1} \, , \label{c2} \eeq
with $C_k^m=\frac{m!}{k!(m-k)!}\,$, 
$~F_{m_1, m_2,\dots,m_r}(q)$ is a polynomial 
of $q$ which is defined in the following, and $\mathcal {P}_N(r)$ denotes the 
set of all partitions (taking care of ordering) of $N$ with length $r$. For 
example, the set $\mathcal {P}_4(2)$ is given by $\big\{ \{3,1\},\{1,3\},
\{2,2\} \big\}$. Let us introduce the partial sums corresponding 
to the partition $\{m_1, m_2 , \dots,m_r \}\in \mathcal {P}_N(r)$ as
\beq M_j=\sum_{i=1}^j m_i, \label{c3} \eeq
where $j \in \{1,2,\dots,r\}$. It may be noted that, $1\leq M_1 <M_2< \dots
< M_{r-1} \leq N-1$ and $M_r=N$. The complementary partial sums 
corresponding to the partition $\{ m_1, m_2,\dots,m_r \}$ are denoted by $M_j$ 
with $j \in \{ r+1,r+2,\dots,N \}$, and they are defined through the relation
\beq \{ M_{r+1}, M_{r+2},\dots, M_N \} \equiv \{1,2,\dots,N \}- \{ 
M_1,M_2,\dots,M_r\}. \label{c4} \eeq
In contrast to the case of partial sums, there exists no natural ordering 
among these complementary partial sums and they can be ordered in an 
arbitrary way. The `energy function' corresponding to a partial sum or 
complementary partial sum is defined as 
\beq \mathcal{E} (M_j)=M_j (N-M_j), \label{c5} \eeq
where $j \in \{1,2,\dots,N \}$.
The polynomial $F_{m_1, m_2,\dots,m_r}(q)$ in \mbox{Eqn. (\ref{c1})} is 
expressed through these energy functions as 
\bea F_{m_1,m_2,\dots,m_r}(q) = q^{\sum\limits_{j=1}^{r-1}\mathcal{E}(M_j)} 
\prod_{j=r+1}^N (1-q^{\mathcal{E}(M_j)}) \, . \label{c6} \eea

It should be noted that $Z_N^{(m|n)}(q)$ in \mbox{Eqn. (\ref{c1})} does not
depend on the parameters $x$ and $y$, which are present in the expression of 
$S_{\langle m_1 , m_2,\dots, m_r \rangle}(x,y)$ in \mbox{Eqn. (\ref{b15})}. 
So, for making a connection with the super Schur polynomials, we introduce
a `generalized' partition function for the $SU(m|n)$ HS spin chain: 
\beq Z_N^{(m|n)} (q;x,y)=\sum_{r=1}^N \sum_{ \{m_1, m_2,\dots,m_r \} \in 
\mathcal{P}_N(r) } \Bigl( \prod_{i=1}^r E_{m_i} (x,y) \Bigr) F_{m_1, m_2,
\dots,m_r}(q), \label{c7} \eeq
$E_{m_i} (x,y)$ being a supersymmetric elementary function which is defined 
in \mbox{Eqn. (\ref{b14})}. Setting all $x_i$'s and $y_i$'s equal to $1$, 
and then equating the coefficients of the same powers of $t$ from both sides 
of \mbox{Eqn. (2.13a)} or \mbox{Eqn. (2.13b)}, 
we find that $$e_k(x)\Bigr\rvert_{x=1}= 
C_k^m, ~~~~ h_{m_i-k}(y)\Bigr\rvert_{y=1}= C_{m_i-k}^{m_i-k+n-1}.$$
Substituting these relations in \mbox{Eqn. (\ref{b14})} 
and comparing it with \mbox{Eqn. (\ref{c2})}, it is easy to see that
\beq E_{m_i}(x,y)\Bigr\rvert_{x=1,y=1}=d^{(m|n)}_{m_i}. \label{c8} \eeq
Consequently, by inserting $x=y=1$ in \mbox{Eqn. (\ref{c7})} and comparing it 
with \mbox{Eqn. (\ref{c1})}, we find that
\beq Z_N^{(m|n)}(q;x,y)\Bigr\rvert_{x=1,y=1}=Z_N^{(m|n)}(q). \label{c9} \eeq

Next, we note that $F_{m_1, m_2,\dots,m_r}(q)$ in \mbox{Eqn. (\ref{c6})} can be
explicitly written in the form of a polynomial as
\beq F_{m_1, m_2,\dots,m_r}(q)=q^{\sum\limits_{j=1}^{r-1} \mathcal{E}(M_j)}
\sum_{\alpha_{r+1}=0}^1 ~\sum_{\alpha_{r+2}=0}^1 \cdots \sum_{\alpha_{N}=0}^1
(-1)^{\, \sum\limits_{i=r+1}^N\alpha_i}
q^{\, \sum\limits_{i=r+1}^N \alpha_i \, \mathcal{E}(M_i)}. \label{c10} \eeq
The lowest power of $q$ in this polynomial is given by
$\sum _{j=1}^{r-1}\mathcal{E}(M_j)$, which is obtained by choosing 
$\alpha_{r+1}=\alpha_{r+2}=\dots=\alpha_N=0$ in the r.h.s. of the above 
equation. On the other hand, for the choice $\alpha_i=1$ when
$i\in \{ l_1,l_2,\dots,l_k \}$ (where $\{l_1,l_2,\dots,l_k\} 
\subseteq \{ r+1, r+2, \dots ,N\}$) and $\alpha_i=0$ when
$i \notin \{ l_1,l_2,\dots,l_k \}$, a term with a higher power of
$q$ given by $\sum_{j=1}^{r-1} \mathcal{E}(M_j)+\sum_{i=1}^k 
\mathcal{E}(M_{l_i})$ will appear in this polynomial. It is easy to check 
that this higher power of $q$ coincides with the lowest power
of $q$ appearing in another polynomial $F_{m_1',m_2',\dots,m_{r+k}'}
(q)$ associated with the partition $\{m_1',m_2',\dots,m_{r+k}' \}
\in \mathcal{P}_N(r+k)$, for which the partial sums form a set given by 
$$\{ M_1,M_2,\dots,M_{r-1},M_r\} \cup \{M_{l_1},M_{l_2},\dots,M_{l_k}\} \, .$$
In this way, any higher power of $q$ appearing in the polynomial
$F_{m_1,m_2,\dots,m_r}(q)$ would coincide with the lowest power of $q$ 
appearing in some other polynomial $F_{m_1',m_2',\dots,m_{r+k}'}(q)$. 
Consequently, the lowest order terms of all possible polynomials like 
$F_{m_1,m_2,\dots,m_r}(q)$ (associated with all possible partitions of $N$) 
form a `complete set', through which $Z_N^{(m|n)}(q;x,y)$ in 
\mbox{Eqn. (\ref{c7})} can be expressed as a polynomial in $q$ as 
\beq Z_N^{(m|n)}(q;x,y)=\sum_{r=1}^N 
\sum_{ \{m_1, m_2,\dots,m_r \} \in \mathcal{P}_N(r)} 
q^{\sum\limits_{j=1}^{r-1} \mathcal{E}(M_j)} 
\tilde{S}_{\l m_1, m_2, \dots,m_r \r}(x,y), \label{c11} \eeq
where $\tilde{S}_{\l m_1, m_2,\dots,m_r \r}(x,y)$ are some unknown functions 
of $x$ and $y$, which will be determined in the following.

Comparing the r.h.s. of \mbox{Eqns. (\ref{c7})} and (\ref{c11}), and also 
using (\ref{c10}), we find that
\bea &&\sum_{r=1}^N \sum_{ \{m_1, m_2, \dots,m_r\} 
\in \mathcal{P}_N(r)} q^{\sum\limits_{j=1}^{r-1}
\mathcal{E}(M_j)} \tilde{S}_{\l m_1, m_2,\dots,m_r \r }(x,y) \nn \\
&& =\sum_{k=1}^N \sum _{ \{ m_1',m_2',\dots,m_k' \} \in \mathcal{P}_N(k)}
\! \left(\prod_{i=1}^k E_{m_i'}(x,y) \! \right) q^{\sum\limits_{j=1}^{k-1} 
\mathcal{E}(M_j')} \! \sum_{\alpha_{k+1}, \cdots , \alpha_N =0}^1
(-1)^{\sum\limits_{i=k+1}^N \! \alpha_i} 
q^{\sum\limits_{i=k+1}^N \! \alpha_i \mathcal{E}(M_i')} , \nn \\
&&~~ \label{c12} \eea
where $M_j'\, $'s denote the partial sums and complementary sums 
corresponding to the partition $ \{m_1', m_2',\dots,m_k'\}$.
Note that corresponding to each partition $ \{m_1', m_2',\dots,m_k'\} \in 
\mathcal{P}_N(k)$, many terms with different powers of $q$ in general appear
in the r.h.s. of the above equation. Let us first 
try to find out the necessary condition for which a partition yields
at least one term with the power of $q$ being given by 
${\sum_{j=1}^{r-1}\mathcal{E}(M_j)}$.
It should be observed that $q^{\sum_{j=1}^{k-1}\mathcal{E}(M_j')}$ is the 
common factor of all terms generated by the partition $\{ m_1',m_2',\dots,
m_k'\}$ in the r.h.s. of \mbox{Eqn. (\ref{c12})}. Consequently, the term 
$q^{\sum_{j=1}^{r-1}\mathcal{E}(M_j)}$ can be generated through the 
partition $ \{ m_1', m_2',\dots, m_k' \} $ only if $k \leq r$ and the
corresponding partial sums satisfy the condition
\beq \{ M_1',M_2', \cdots,M_k' \}\subseteq \{ M_1,M_2, \cdots,M_r \}.
\label{c13} \eeq
Hence, we can write these partial sums as
\beq M_i'=M_{L_i}, \label{c14} \eeq
where $i \in \{ 1,2,\dots,k \}$, and the indices $L_1,L_2,\dots,L_k$ satisfy
the condition
\beq 1 \leq L_1 <L_2 \dots <L_k=r . \label{c15} \eeq
Let us consider another set of indices like $L_{k+1},L_{k+2},\dots,L_r$,
and define the corresponding set as 
\beq
\{ L_{k+1},L_{k+2},\dots,L_r \} \equiv \{1,2,\dots,r\}-\{L_1,L_2,\dots,L_k \}.
\label{c16} \eeq
Using \mbox{Eqns. (\ref{c14})} and (\ref{c16}) we obtain
\bea \{ M_1,M_2,\dots,M_r\}-\{M_1',M_2',\dots,M_k'\}\hskip -.5cm &&= 
\{ M_1,M_2,\dots,M_r\}-\{M_{L_1},M_{L_2},\dots,M_{L_k}\}\nn\\
&&= \{M_{L_{k+1}},M_{L_{k+2}},\dots,M_{L_{r}}\}. \label{c17} \eea
With the help of the embedding condition 
$\{ M_1',M_2', \cdots,M_k' \}\subseteq \{ M_1,M_2, \cdots,M_r \} 
\subseteq \{1,2, \cdots , N \}$, the set of complementary partial sums 
associated with the partition $\{m_1',m_2',\dots,m_k' \}$ can be written as 
\bea && \{M_{k+1}',M_{k+2}',\dots,M_{N}' \} \nn \\
&& = \{ 1,2,\dots,N \}- \{M_1',M_2',\dots,M_k' \} \nn \\
&& = \big( \{ 1,2,\dots,N \}-\{M_1,M_2,\dots,M_r\} \big) \cup \big( 
\{M_1,M_2,\dots,M_r \}-\{M_1',M_2',\dots,M_k' \}\big) . \nn \\
&& \label{c18} \eea
Using the relation (\ref{c18}) along with (\ref{c4}) and (\ref{c17}), 
we find that 
\beq \{M_{k+1}',M_{k+2}',\dots,M_{N}' \} =\{M_{r+1},M_{r+2},\dots,M_N\} \cup 
\{M_{L_{k+1}},M_{L_{k+2}},\dots,M_{L_r}\}. \label{c19} \eeq
Consequently, we can express the complementary partial sums associated with 
the partition $\{m_1',m_2',\dots,m_k'\}$ as 
\beq M_j'=M_{L_j}, \label{c20} \eeq
where the $L_j$'s are defined through \mbox{Eqn. (\ref{c16})} when $j\in 
\{k+1,k+2,
\cdots , r\}$ (the ordering of indices $L_{k+1},L_{k+2},\dots,L_r$ is not 
important for our purpose), and $L_j=j$ when $j \in \{r+1,r+2,\dots,N \}$.

With the help of \mbox{Eqns. (\ref{c17})} and (\ref{c20}) we find that, for the
choice of summation variables like $\alpha_{k+1}=\alpha_{k+2}=\dots=\alpha_r
=1$ and $\alpha_{r+1}=\alpha_{r+2}=\dots=\alpha_N=0$, one term with the power 
of $q$ given by ${\sum_{j=1}^{r-1}\mathcal{E}(M_j)}$ is generated through the 
partition $\{m_1',m_2',\dots,m_k'\}$ in the r.h.s. of \mbox{Eqn. (\ref{c12})}. 
Moreover, the coefficient of $q^{\sum_{j=1}^{r-1} \mathcal{E}(M_j)}$ in 
the above mentioned term is obtained as 
\beq
\mathcal{C}_{\{m_1',m_2',\dots,m_k'\}}=(-1)^{r-k} \prod_{i=1}^k E_{m_i'}(x,y) 
= (-1)^{r-k} \prod_{i=1}^k E_{M_i'-M_{i-1}'}(x,y) \, , \label{c21} \eeq
where it is assumed that $M_0'=0$. Using \mbox{Eqn. (\ref{c14})}, these 
coefficients can also be written as 
\beq \mathcal{C}_{\{m_1',m_2',\dots,m_k'\}}
=(-1)^{r-k} \prod_{i=1}^k E_{M_{L_i}-M_{L_{i-1}}}(x,y)\, , \label{c22} \eeq
where the indices $L_1,L_2,\dots,L_k$ satisfy the condition (\ref{c15}) 
and it is assumed that $M_{L_0}=0$. 

{}From the above discussion it is evident that, \mbox{Eqn. (\ref{c13})} 
represents 
not only the \emph{necessary} but also the \emph{sufficient} condition for 
which the partition $\{m_1',m_2',\dots,m_k'\}$ yields one term with the power 
of $q$ being given by ${\sum_{j=1}^{r-1}\mathcal{E}(M_j)}$. By summing up the 
coefficients of $q^{\sum_{j=1}^{r-1}\mathcal{E}(M_j)}$ associated with all 
such partitions, we can determine $\tilde{S}_{\l m_1, m_2, \dots,m_r \r }(x,y)$
appearing in \mbox{Eqn. (\ref{c12})}. 
Thus, by using \mbox{Eqn. (\ref{c22})}, we find that 
\beq \tilde{S}_{\l m_1,m_2,\dots,m_r \r}(x,y) =\sum_{k=1}^r ~ \sum_{1 \leq L_1
<\dots<L_k=r} (-1)^{r-k} \prod_{i=1}^k E_{M_{L_i}-M_{L_{i-1}}}(x,y) \, .
\label{c23} \eeq
Since the $L_i$'s appearing in the above equation satisfy the condition 
(\ref{c15}), they can be written as 
\beq L_i=\sum_{s=1}^i \ell_s \, , \label{c24} \eeq
where the $\ell_s$'s are $k$ number of positive integers such that 
$\{\ell_1,\ell_2,\dots,\ell_k\} \in \mathcal{P}_r(k)$. Consequently, 
$\tilde{S}_{\l m_1,m_2,\dots,m_r \r }(x,y)$ in \mbox{Eqn. (\ref{c23})} can 
also be expressed in the form
\beq \tilde{S}_{ \l m_1,m_2,\dots,m_r \r}(x,y)=\sum_{k=1}^r 
\sum_{\{\ell_1,\ell_2,\dots,\ell_k\} \in \mathcal{P}_r(k)}
(-1)^{r-k}\prod_{i=1}^kE_{M_{L_i}-M_{L_{i-1}}}(x,y) \, , \label{c25} \eeq
where the $L_i$'s are related to the $\ell_s$'s through \mbox{Eqn. 
(\ref{c24})}.

Next, we find that $\tilde{S}_{ \l m_1,m_2,\dots,m_r \r }(x,y)$ given by 
\mbox{Eqn. (\ref{c25})} satisfies the following recursion relation:
\beq \tilde{S}_{\l m_1, m_2, 
\dots,m_r \r}(x,y) =\sum_{s=1}^{r} (-1)^{s+1} E_{m_r+m_{r-1}+
\dots+m_{r-s+1}}(x,y) \cdot \tilde{S}_{\l m_1, m_2,\dots,m_{r-s} \r }(x,y)\, ,
\label{c26} \eeq
where $\tilde{S}_{\l 0 \r}(x,y)=1$. The derivation of this recursion relation 
is presented in Appendix A. It is interesting to observe that, the above 
recursion relation is exactly the same in form as the recursion relation 
(\ref{b16}), which is satisfied by the super Schur polynomials associated 
with the border strips. Consequently, the function 
$\tilde{S}_{\l m_1, m_2, \dots, m_r \r}(x,y)$ coincides with the super Schur 
polynomial $S_{ \l m_1,m_2,\dots,m_r \r} (x,y)$. Thus, \mbox{Eqn. (\ref{c25})}
gives us a novel combinatorial formula for the super Schur polynomial 
corresponding to the border strip $\l m_1,m_2,\dots,m_r \r $, which is usually
defined through \mbox{Eqn. (\ref{b15})}. Substituting $\tilde
{S}_{ \l m_1,m_2,\dots,m_r \r }(x,y)$ by 
$S_{\l m_1,m_2,\dots,m_r \r}(x,y)$ in \mbox{Eqn. (\ref{c11})}, 
we can express the generalized partition function 
of the $SU(m|n)$ HS spin through the super Schur polynomials as 
\beq Z_N^{(m|n)}(q;x,y)=\sum_{r=1}^N 
\sum_{ \{m_1,m_2, \dots,m_r \} \in \mathcal{P}_N(r)} 
q^{\sum\limits_{j=1}^{r-1} \mathcal{E}(M_j)} 
S_{\l m_1, m_2, \dots,m_r \r }(x,y) \, . \label{c27} \eeq
In the limit $x=y=1$, this generalized partition function clearly reduces to 
the standard partition function $Z_N^{(m|n)}(q)$ given in 
\mbox{Eqn. (\ref{c1})}.

It should be noted that, the partition function of the Polychronakos model 
can also written in a form similar to (\ref{c27}) by using the corresponding 
energy function [32,22]. This is due to the fact that both models, the HS 
model and the Polychronakos model, share the same Yangian symmetry.
See \mbox{Ref. 35} for the conformal field theoretic construction of
conserved quantities leading to the Yangian symmetry.

\vskip 1cm 
\noi \section{Duality relation for supersymmetric HS spin chain}
\renewcommand{\theequation}{4.{\arabic{equation}}}
\setcounter{equation}{0}

In this section, our aim is to give an analytical proof for
the duality relation (\ref{a2}) involving the partition functions of the 
$SU(m|n)$ and $SU(n|m)$ HS spin chains. A central role in this proof will be 
played by the duality relation (\ref{b12}) for the super Schur polynomials.
Let us assume that $\lambda/\mu$ in (\ref{b12}) represents a border strip: 
$\lambda/\mu \equiv \l m_1, m_2, \dots,m_r \r$, where $\{ m_1, m_2, \dots,
m_r \} \in \mathcal{P}_N(r)$, and denote the conjugate of this border strip
as $\l m_1, m_2, \dots,m_r \r'$. Applying the rule 
for obtaining a conjugate diagram to \mbox{Fig. 1}, we find that 
\vskip .11cm
\begin{figure}[h]
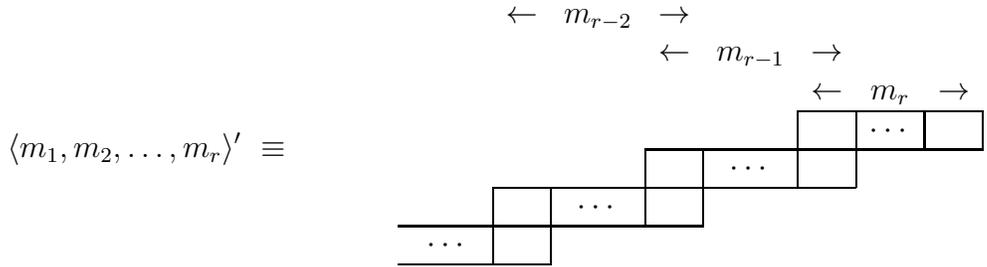

$$
\l m_1, m_2, \dots,m_r \r' 
~\equiv~~~~~~~~~~
\begin{array}{cccccccc}
{} & \leftarrow & m_{r-2} & \rightarrow & \mc{4}{c}{} \\
\mc{3}{c}{} & \leftarrow & m_{r-1} & \rightarrow & \mc{2}{c}{} \\
\mc{5}{c}{} & \leftarrow & m_{r} & \rightarrow \\ 
\cline{6-8} \mc{5}{c|}{} & &\mc{1}{|c|}{\cdots} &\mc{1}{c|}{} \\
\cline{4-8} \mc{3}{c|}{} & &\mc{1}{|c|}{\cdots} &\mc{1}{c|}{} \\
\cline{2-6} \mc{1}{c|}{} & &\mc{1}{|c|}{\cdots} &\mc{1}{c|}{} \\
\cline{1-4} \mc{1}{c|}{~\cdots ~ } &\mc{1}{c|}{} &\mc{6}{c}{} \\
\cline{1-2} \\
\end{array} \, .$$
\caption{Shape of the border strip conjugate to $\l m_1,m_2,\dots,m_r \r$}
\end{figure} 

It may be observed that, the vertical length of the border strip $\l m_1, 
m_2, \dots,m_r \r$ drawn in \mbox{Fig. 1} is given by $(N-r+1)$. Since this 
vertical length coincides with the horizontal length of the conjugate 
border strip $\l m_1, m_2, \dots,m_r \r'$ (as drawn in \mbox{Fig. 2}), 
it can also be expressed as 
$$\l m_1, m_2, \dots,m_r \r' \equiv \l m_1', m_2', \dots , m_{N-r+1}' \r ,$$
where $m_j'$'s are some functions of $m_j$'s and $\{ m_1', 
m_2', \dots , m_{N-r+1}' \}\in \mathcal{P}_N(N-r+1)$. A relation between 
$m_j$'s and $m_j'$'s will be established in the following by exploiting the 
motif representations corresponding to the border strips. 

Using the rule for mapping a motif to a border strip as discussed in Sec. 2, 
and observing \mbox{Fig. 1}, it is easy to check that 
\beq (\,\underbrace{1,\dots,1}_{m_1-1}\, , 0, \underbrace{1,\dots,1}_{m_2-1},0,
\dots \dots,0,\underbrace{1,\dots,1}_{m_r-1}\,) ~\Longrightarrow ~ \l m_1, 
m_2, \dots,m_r \r \, . \label{d1} \eeq
Let us denote the set formed by the positions of $0$'s (resp. $1$'s) in the 
motif associated with the border strip $ \l m_1, m_2, \dots,m_r \r \, $ as 
$Q_{\l m_1, m_2, \dots,m_r\r } (0)$ (resp. $Q_{\l m_1, m_2, \dots,m_r \r} 
(1)$). For example, since the motif $(10110)$ leads to the border strip
$\l 2,3,1 \r$, $Q_{\l 2,3,1 \r}(0)= \{2,5\}$ and $Q_{\l 2,3,1 \r}(1)= 
\{1,3,4\}$. Observing the l.h.s. of (\ref{d1}) we find that, the set $Q_{\l 
m_1, m_2, \dots,m_r\r } (0)$ can be expressed through the partial sums 
corresponding to the partition $\{ m_1, m_2, \dots,m_r \}$ as 
\beq Q_{ \l m_1, m_2, \dots,m_r \r}(0) = \{ M_1, M_2, \cdots , M_{r-1} \} \, .
\label{d2} \eeq
Using (\ref{d2}) along with the relation $$Q_{ \l m_1, m_2, \dots,m_r \r }(1)
= \{1,2, \cdots , N-1\} - Q_{ \l m_1, m_2, \dots,m_r \r}(0) \, ,$$ it is easy 
to express the set $ Q_{\l m_1, m_2, \dots,m_r\r } (1) $ through the 
complementary partial sums corresponding to the partition $\{ m_1, m_2, 
\dots, m_r \}$ as 
\beq 
Q_{ \l m_1, m_2, \dots,m_r \r }(1) = \{ M_{r+1}, M_{r+2}, \cdots , M_N \} \, .
\label{d3} \eeq
Applying the rule for mapping a motif to a border strip and observing
\mbox{Fig. 2}, it is easy to check that
\beq
(\,\underbrace{0,\dots,0}_{m_r-1},1, \underbrace{0, \dots,0}_{m_{r-1}-1},1,
\dots \dots ,1, \underbrace{0,\dots,0}_{m_1-1}\,) ~ \Longrightarrow ~ 
\l m_1',m_2',\dots,m_{N-r+1}' \r \, . \label{d4} \eeq
Comparing the l.h.s. of (\ref{d1}) and (\ref{d4}) we find interestingly that
the conjugate motif can be obtained from the original motif by applying the 
following two rules:
\begin{itemize}
\item Replacing $0$'s with $1$'s and vice versa, 
\item Rewriting all binary digits in the reverse order. 
\end{itemize}
For example, the conjugate of the motif $(10110)$ is obtained as 
$(10110) \rightarrow (01001) \rightarrow (10010)$. Using the above 
mentioned rules along with \mbox{Eqn. (\ref{d3})}, it is found that 
\beq Q_{ \l m_1', m_2', \dots,m_{N-r+1}' \r }(0)
= \{ N-M_{r+1}, N-M_{r+2}, \cdots , N-M_N \} \, . \label{d5} \eeq
On the other hand, using \mbox{Eqn. (\ref{d2})} for the case of border strip 
$\l m_1', m_2', \dots , m_{N-r+1}' \r$, we obtain
\beq Q_{ \l m_1', m_2', \dots,m_{N-r+1}' \r }(0) 
= \{ M_1', M_2', \cdots , M_{N-r}' \} \, , \label{d6} \eeq
where $M_i'$'s denote the first $(N-r)$ number of partial sums corresponding 
to the partition $\{ m_1', m_2', \dots , m_{N-r+1}' \}$. Comparing the r.h.s. 
of (\ref{d5}) with (\ref{d6}) we find that 
\beq \{ M_1', M_2', \cdots , M_{N-r}' \} = \{ N-M_{r+1}, N-M_{r+2}, \cdots ,
N-M_N \} \, . \label{d7} \eeq
This relation between the partial sums associated with the conjugate border 
strip $\l m_1', m_2', \dots , m_{N-r+1}' \r$ and complementary partial sums 
associated with the border strip $\l m_1, m_2, \dots , m_r \r$ will be used 
shortly for proving the duality relation (\ref{a2}).

With the help of \mbox{Eqn. (\ref{b12})}, we express the generalized partition 
function (\ref{c27}) as
\beq Z_N^{(m|n)}(q;x,y)
=\sum_{r=1}^N \sum_{\{ m_1,m_2,\dots,m_r \} \in \mathcal{P}_N(r)} 
q^{\sum\limits_{j=1}^{r-1} \mathcal{E}(M_j)} 
S_{\l m_1', m_2',\dots,m_{N-r+1}'\r}(y,x) \, . \label{d8} \eeq
Due to \mbox{Eqn. (\ref{c4})}, it follows that $\{M_1, M_2, \dots , M_N\} = 
\{1,2, \dots N \}$. So we obtain $$\sum_{j=1}^N \mathcal{E}(M_j)= \sum_{j=1}^N
j(N-j) = \frac{N(N^2-1)}{6}.$$ Using this relation along with the fact that 
$\mathcal{E}(M_r) =0$, one can rewrite \mbox{Eqn. (\ref{d8})} as 
\beq Z_N^{(m|n)}(q;x,y) =q^{\frac{N(N^2-1)}{6}}
\sum_{r=1}^N \sum_{\{ m_1, m_2,\dots,m_r \} \in \mathcal{P}_N(r)} 
q^{- \! \sum\limits_{j=r+1}^{N} \mathcal{E}(M_j)} 
S_{\l m_1', m_2', \dots,m_{N-r+1}'\r}(y,x) \, . \label{d9} \eeq
{}From the definition of $\mathcal{E}(M_j)$ 
in \mbox{Eqn. (\ref{c5})}, it is evident 
that $$\mathcal{E}(M_j)=\mathcal{E}(N-M_j).$$ Applying this relation along 
with \mbox{Eqn. (\ref{d7})}, we obtain $$\sum_{j=r+1}^N \mathcal{E}(M_j)=
\sum_{j=r+1}^N \mathcal{E}(N-M_j) =\sum_{\ell=1}^{N-r} \mathcal{E}(M_\ell').$$
Using the above relation and rearranging the summation variables,
$Z_N^{(m|n)}(q;x,y)$ in \mbox{Eqn. (\ref{d9})} may be written as 
\bea Z_N^{(m|n)}(q;x,y) 
\hskip -.2cm &=& \hskip -.2cm q^{\frac{N(N^2-1)}{6}}
\sum_{r=1}^N \, \sum_{\{ m_1,m_2,\dots,m_r \} \in \mathcal{P}_N(r)} 
q^{-\sum\limits_{\ell=1}^{N-r} \mathcal{E}(M_{\ell}')} 
S_{\l m_1',m_2',\dots,m_{N-r+1}'\r } (y,x) \nn \\
\hskip -.2cm &=& \hskip -.2cm q^{\frac{N(N^2-1)}{6}} 
\sum_{s=1}^N \, \sum_{ \{ m_1', m_2',\dots,m_{s}' \} 
\in \mathcal{P}_N(s)} q^{-\sum\limits_{\ell=1}^{s-1} \mathcal{E}(M_{\ell}')}
S_{\l m_1',m_2',\dots,m_{s}'\r }(y,x) \, , ~~ \label{d10} \eea
where we have used the notation $s\equiv N-r+1$. With the help of 
(\ref{c27}), \mbox{Eqn. (\ref{d10})} is finally expressed as 
\beq Z_N^{(m|n)}(q;x,y) = q^{\frac{N(N^2-1)}{6}} Z^{(n|m)}_N(q^{-1};y,x).
\label{d11} \eeq
Thus we are able to prove a duality relation between the generalized 
partition functions of the $SU(m|n)$ and $SU(n|m)$ HS spin chains under the 
exchange of bosonic and fermionic degrees of freedom. In the limit $x=y=1$, 
this duality relation reduces to the duality relation (\ref{a2}) for the 
partition functions of supersymmetric HS spin chains.
\vskip 1 cm 

\noi \section{Duality in spin models with global $SU(m|n)$ symmetry}
\renewcommand{\theequation}{5.{\arabic{equation}}}
\setcounter{equation}{0}

The super Yangian symmetry of the $SU(m|n)$ HS spin chain has clearly played 
a key role in the previous section for establishing the duality relation 
(\ref{a2}). However, such a Yangian symmetry is found to exist only in very 
few quantum integrable spin chains. So it is natural to ask whether 
nonintegrable spin models can also exhibit duality relation under the 
exchange of bosonic and fermionic spin degrees of freedom. In the following, 
we shall try to answer this question by using a rather different approach. 

Let us consider a Hamiltonian of the form 
\beq \mathcal{H}^{(m|n)} = \omega_0 + \sum_{1\leq j < k \leq N} \omega_{jk} \,
\hat{P}_{jk}^{(m|n)} \, , \label{e1} \eeq
where $\omega_0$ and $\omega_{jk}$'s are arbitrary constant parameters and 
$\hat{P}_{jk}^{(m|n)}$ is the supersymmetric exchange operator defined in 
\mbox{Eqn. (\ref{b3})}. Similar to the case of the $SU(m|n)$ supersymmetric HS
spin chain, the action of Hamiltonian (\ref{e1}) is restricted to the state 
vectors which satisfy the condition (\ref{b2}). Due to this condition, the 
supersymmetric exchange operator $\hat{P}_{jk}^{(m|n)}$ becomes equivalent to 
an `anyon like' representation of the permutation algebra [20]. The vector 
space corresponding to this anyon like representation of the permutation 
algebra is a direct product of $N$ number of $(m+n)$-dimensional spin spaces,
and it contains orthonormal basis vectors like $\v \alpha_1 \alpha_2 \cdots
\alpha_j \cdots \alpha_N \r_{m,n} \,$, 
where $\alpha_j \in \{ 1,2,\dots,m+n \}$. 
Let us denote this space as $V_{(m|n)}$. For the sake of convenience, we 
assign a `parity' $p(\alpha_j)$ to each spin component $\alpha_j$. Moreover, 
we call $\alpha_j$ a `bosonic' spin with $p(\alpha_j)=0$ when $\alpha_j \in 
\{ 1,2,\dots,m \}$ and a `fermionic' spin with $p(\alpha_j)=1$ when 
$\alpha_j \in \{m+1,m+2, \dots,m+n \}$. The symbol $\rho_{jk}^{\alpha}(f)$ 
denotes the total number of fermionic spins lying in between the $j$-th and 
$k$-th lattice sites for the case of state vector $\v \alpha_1 \alpha_2 
\cdots \alpha_j \cdots \alpha_k \cdots \alpha_N \r_{m,n} \, $:
\beq \rho_{jk}^\alpha(f) \equiv \sum_{\ell = j+1}^{k-1} p(\alpha_\ell) \, .
\label{e2} \eeq
If we define an anyon like representation $\tilde{P}_{jk}^{(m|n)}$ 
on the space $V_{(m|n)}$ as 
\bea && \tilde{P}_{jk}^{(m|n)} \v \alpha_1 \alpha_2 \cdots \alpha_j \cdots 
\alpha_k \cdots \alpha_N \r_{m,n} \nn \\
&& ~~~~~~~~~= (-1)^{ p(\alpha_j)p(\alpha_k) + \, \left\{ p(\alpha_j) + 
p(\alpha_k) \right\} \, \rho_{jk}^\alpha(f)} \, \v \alpha_1 \alpha_2 \cdots 
\alpha_k \cdots \alpha_j \cdots \alpha_N \r_{m,n} ~ , \label{e3} \eea
that will be equivalent to the supersymmetric exchange operator
$\hat{P}_{jk}^{(m|n)}$ given in \mbox{Eqn. (\ref{b3})} [20,23]. 
The relation (\ref{e3}) implies that the exchange of two bosonic (resp. 
fermionic) spins produces a phase factor of $ 1 \, (\text{resp.} -1)$ 
irrespective of the nature of the spins situated in between the $j$-th and 
$k$-th lattice sites. However, if we exchange one bosonic spin with one 
fermionic spin, then the phase factor becomes $1~(\text{resp.} -1) $ if there 
exist even (resp. odd) number of fermionic spins situated
in between the $j$-th and $k$-th lattice sites. Due to the above mentioned 
equivalence between the supersymmetric exchange operator 
$\hat{P}_{jk}^{(m|n)}$ and the anyon like representation 
$\tilde{P}_{jk}^{(m|n)}$, the Hamiltonian $\mathcal{H}^{(m|n)}$ in 
\mbox{Eqn. (\ref{e1})} is equivalently expressed as 
\beq \mathcal{H}^{(m|n)} = \omega_0 + \sum_{1\leq j < k \leq N} \omega_{jk} \,
\tilde{P}_{jk}^{(m|n)} \, . \label{e4} \eeq

Let us now define another set of orthonormal basis vectors for the space 
$V_{(m|n)}$ by multiplying the states like $\v \alpha_1 \alpha_2 \cdots 
\alpha_j \cdots \alpha_N \r_{m,n} $ through a phase factor, which takes the 
value $+1$ (resp. $-1$) when the total number of fermionic spins sitting 
on all odd numbered lattices sites is even (resp. odd): 
\beq \v \alpha_1 \alpha_2 \cdots \alpha_j \cdots \alpha_N \r_{m,n}^* 
= (-1)^{\sum\limits_{\ell=1}^N \ell \, p(\alpha_\ell)} \,
\v \alpha_1 \alpha_2 \cdots \alpha_j 
\cdots \alpha_N \r_{m,n} \, . \label{e5} \eeq
By using \mbox{Eqns. (\ref{e3})} and (\ref{e5}) we find that the action of 
$\tilde{P}_{jk}^{(m|n)}$ on these new basis vectors is given by 
\bea
&& \tilde{P}_{jk}^{(m|n)} \v \alpha_1 \alpha_2 \cdots \alpha_j \cdots \alpha_k
\cdots \alpha_N \r_{m,n}^* \nn \\
&& ~=(-1)^{\sum\limits_{\ell=1}^N \ell \, p(\alpha_\ell)} 
\tilde{P}_{jk}^{(m|n)} \v \alpha_1 \alpha_2 \cdots \alpha_j \cdots \alpha_k 
\cdots \alpha_N \r_{m,n} \nn \\
&& ~=(-1)^{\sum\limits_{\ell=1}^N \ell\, p(\alpha_\ell)}
(-1)^{p(\alpha_j) p(\alpha_k) + \, \left\{p(\alpha_j) + p(\alpha_k) 
\right\} \, \rho_{jk}^\alpha(f)} \, \v \alpha_1 \alpha_2 \cdots \alpha_k 
\cdots \alpha_j \cdots \alpha_N \r_{m,n} \nn \\
&& ~= (-1)^{p(\alpha_j) p(\alpha_k) + \, \left\{p(\alpha_j) + p(\alpha_k) 
\right\} \, \left( \rho_{jk}^\alpha(f)+j+k \right)} \, 
\v \alpha_1 \alpha_2 \cdots \alpha_k \cdots \alpha_j \cdots 
\alpha_N \r_{m,n}^* ~. \label{e6} \eea

Let us now consider the vector space $V_{(n|m)}$, which might be spanned 
through orthonormal basis vectors like $\v \beta_1 \beta_2 \cdots \beta_j 
\cdots \beta_N \r_{n,m} \,$, where $\beta_j \in \{ 1,2,\dots,m+n \}$. In 
this case, we call $\beta_j$ a `bosonic' spin with $p(\beta_j)=0$ when 
$\beta_j \in \{ 1,2,\dots,n \} $ and a `fermionic' spin with $p(\beta_j)=1$ 
when $\beta_j \in \{n+1,n+2,\dots,n+m \} $. In analogy with 
\mbox{Eqn. (\ref{e3})}, we can express 
the action of $\tilde{P}_{jk}^{(n|m)}$ on the space $V_{(n|m)}$ as 
\bea
&& \tilde{P}_{jk}^{(n|m)} \, \v \beta_1 \beta_2 \cdots \beta_j \cdots \beta_k 
\cdots \beta_N \r_{n,m} \nn \\
&&~~~~~~~~~=
(-1)^{ p(\beta_j)p(\beta_k) + \, \left\{ p(\beta_j) + p(\beta_k) 
\right\} \, \rho_{jk}^\beta(f)} \, \v \beta_1 \beta_2 \cdots \beta_k \cdots 
\beta_j \cdots \beta_N \r_{n,m} \, , \label{e7} \eea
where $ \rho_{jk}^\beta(f)$ denotes the total number of fermionic spins lying 
in between the $j$-th and $k$-th lattice sites in the case of the state vector
$\v \beta_1 \beta_2 \cdots \beta_j \cdots \beta_k \cdots \beta_N \r_{n,m} \,$:
\beq \rho_{jk}^\beta(f) \equiv \sum_{\ell = j+1}^{k-1} p(\beta_\ell) \, .
\label{e8} \eeq
Let $U$ be a permutation of the set $\{ 1,2,\dots , m+n \}$, which satisfies 
the relations
\bea
&& \{ \, U(1),U(2), \dots ,U(m) \, \} = \{ n+1, n+2, \dots , n+m \} \, , \nn \\
&& \{ \, U(m+1),U(m+2), \dots ,U(m+n) \, \} = \{ 1, 2, \dots , n \} \, . 
\label{e9} \eea
With the help of this permutation, we define an unitary operator 
($\mathcal{U}$) which maps the vectors of $V_{(m|n)}$
to the vectors of $V_{(n|m)}$ as 
\beq \mathcal{U} \v \alpha_1 \alpha_2 \cdots \alpha_j
\cdots \alpha_N \r_{m,n}^* = 
\v \beta_1 \beta_2 \cdots \beta_j \cdots \beta_N \r_{n,m} \, , \label{e10} \eeq
where $\beta_j\equiv U(\alpha_j)$. If $\alpha_j$ represents a bosonic (resp. 
fermionic) spin in the space $V_{(m|n)}$, then, due to \mbox{Eqn. (\ref{e9})}, 
$\beta_j$ would represent a fermionic (resp. bosonic) spin in the space 
$V_{(n|m)}$. Hence we can write 
\beq p(\beta_j) = 1 - p(\alpha_j) \, . \label{e11} \eeq
Using the relations (\ref{e2}), (\ref{e8}) and (\ref{e11}), we also obtain 
\beq \rho_{jk}^\beta(f) =(j-k-1)-\rho_{jk}^\alpha(f) \, .
\label{e12} \eeq
With the help of \mbox{Eqns. (\ref{e10})}, (\ref{e11}) and (\ref{e12}), one can 
easily express equation (\ref{e7}) in the form
\bea && \mathcal{U^\dagger} \, \tilde{P}_{jk}^{(n|m)} \, \mathcal{U} \, \v 
\alpha_1 \alpha_2 \cdots \alpha_j \cdots \alpha_k \cdots \alpha_N 
\r_{m,n}^* \nn \\
&& ~= -(-1)^{ p(\alpha_j)p(\alpha_k) + \, \left\{ p(\alpha_j) + p(\alpha_k)
\right\} \, \left(\rho_{jk}^\alpha(f)+j+k \right)} \,
\v \alpha_1 \alpha_2 \cdots \alpha_k \cdots \alpha_j \cdots \alpha_N 
\r_{m,n}^* \, . ~~~~~ \label{e13} \eea
Comparison of \mbox{Eqn. (\ref{e6})} with \mbox{Eqn. (\ref{e13})} implies that
\beq \mathcal{U} \tilde{P}_{jk}^{(m|n)}\mathcal{U}^\dagger = - 
\tilde{P}_{jk}^{(n|m)} \, , \label{e14} \eeq
for all possible values of $j$ and $k$. Using \mbox{Eqns. (\ref{e4})} and 
(\ref{e14}), we find that 
\beq \mathcal{U} \mathcal{H}^{(m|n)}\mathcal{U}^\dagger = 2\omega_0 - 
\mathcal{H}^{(n|m)} \, . \label{e15} \eeq
Hence the spectrum of the Hamiltonian $\mathcal{H}^{(m|n)} $ can be obtained 
by inverting the spectrum of its dual Hamiltonian $\mathcal{H}^{(n|m)}$ and 
giving this inverted spectrum an overall shift of amount $2\omega_0$. If we 
define the partition function corresponding to the Hamiltonian 
$\mathcal{H}^{(m|n)}$ as $\mathcal{Z}_N^{(m|n)}(q)\equiv tr \big(
q^{\mathcal{H}^{(m|n)} }\big)$, then by using \mbox{Eqn. (\ref{e15})} we 
obtain a duality relation at the level of the partition function as 
\beq
\mathcal{Z}_N^{(m|n)}(q) = q^{2\omega_0} \, \mathcal{Z}_N^{(n|m)}(q^{-1}) \, .
\label{e16} \eeq
It should be observed that, while proving the above duality relation,
we have kept the coupling constants $\omega_0$ and $\omega_{jk}$
in the Hamiltonian $\mathcal{H}^{(m|n)}$ (\ref{e1}) as completely free 
parameters. By properly choosing these free parameters, one can generate 
many quantum integrable models like $SU(m|n)$ supersymmetric versions of 
the Haldane-Shastry spin chain, the Polychronakos spin 
chain, and the isotropic Heisenberg spin chain. For example, by choosing 
$\omega_0 = \sum_{1\leq j <k \leq N} \frac{1}{\sin^2(\xi_j - \xi_k )} = 
\frac{N(N^2-1)}{12}$ and $\omega_{jk} = \frac{1}{\sin^2(\xi_j - \xi_k )}$, 
with $\xi_j = \frac{j\pi}{N}$, in \mbox{Eqn. (\ref{e1})}, we recover 
the Hamiltonian (\ref{a1}) of the $SU(m|n)$ supersymmetric HS spin chain. 
The partition functions for all of the above mentioned quantum integrable 
models will naturally satisfy the duality relation (\ref{e16}). 
However, since the spin chain Hamiltonian (\ref{e1}) is nonintegrable for 
generic choice of $\omega_{jk}$, it is obvious that integrability or quantum 
group symmetry is not a necessary requirement for the existence of the
duality relation (\ref{e16}). 

By using the (anti-)commutation relations (\ref{b1}), 
it is easy to verify that the supersymmetric exchange operator
$\hat{P}_{jk}^{(m|n)}$ (\ref{b3}) commutes with the set of operators 
$ Q_0^{\alpha \beta} $ given in \mbox{Eqn. (2.10a)}. Therefore, 
the spin chain Hamiltonian $\mathcal{H}^{(m|n)}$ (\ref{e1}) 
also commutes with the set of operators $ Q_0^{\alpha \beta} $. Since 
commutation relations among the $Q_0^{\alpha \beta}$'s generate the 
$SU(m|n)$ algebra, it is clear that $\mathcal{H}^{(m|n)}$ has a global 
$SU(m|n)$ symmetry for any value of the parameters $\omega_0$ and 
$\omega_{jk}$. It may be noted that $\mathcal{H}^{(m|n)}$ in \mbox{Eqn. 
(\ref{e1})} depends linearly on the supersymmetric exchange operators 
$\hat{P}_{jk}^{(m|n)}$. One can also construct more general spin chain
Hamiltonians with a global $SU(m|n)$ symmetry by including the products of 
different exchange operators (corresponding to different lattice sites)
with arbitrary coefficients. It 
is easy to see that such a Hamiltonian would satisfy the duality relation 
(\ref{e16}), provided we construct the dual Hamiltonian $\mathcal{H}^{(n|m)}$ 
from the original Hamiltonian $\mathcal{H}^{(m|n)}$ by keeping $\omega_0$ as 
well as all coupling constants associated with the products of odd numbers of 
exchange operators unchanged, reversing the sign of all coupling constants 
associated with the products of even numbers of exchange operators, and 
finally replacing $\hat{P}_{jk}^{(m|n)}$ by $\hat{P}_{jk}^{(n|m)}$. Since any 
quantum integrable or nonintegrable spin chain with a global $SU(m|n)$ 
symmetry can be expressed as a polynomial function of the exchange operators 
$\hat{P}_{jk}^{(m|n)}$, it is evident that the partition functions associated 
with such spin chains would satisfy the duality relation (\ref{e16}). 

\noi \section{Concluding remarks}

We have provided here an analytical proof for the boson-fermion duality 
relation in the case of the $SU(m|n)$ supersymmetric HS spin chain. To this 
end, we utilize the $Y(gl(m|n))$ super Yangian symmetry of the $SU(m|n)$ 
HS spin chain in a crucial way. At first, we define a generalized partition 
function which reduces to the usual partition function of this spin chain
in some limit of the related parameters. Subsequently, we express this 
generalized partition function in terms of the super Schur polynomials
associated with border strips, which label a family
of irreducible representations of the $Y(gl(m|n))$
quantum group. It is well known that such super Schur polynomials satisfy 
a duality relation under the exchange of bosonic and fermionic variables. 
Using this duality relation, we finally derive the boson-fermion duality 
(\ref{a2}) for the partition function of the $SU(m|n)$ HS spin chain.
Apart from leading to a proof for this duality relation, 
our expression (\ref{c27}) for the generalized partition function 
of the $SU(m|n)$ HS spin chain through the super Schur polynomials
might be useful in future for finding various correlation functions.

As a byproduct of our investigation of the duality relation of the $SU(m|n)$ 
HS spin chain, we obtain some results which are interesting from the 
mathematical point of view. For example, while expressing the partition 
function of the $SU(m|n)$ HS spin chain in terms of the super Schur 
polynomials, we find a novel combinatorial formula for the super Schur 
polynomials. Furthermore, while studying the transformation of a border 
strip under the conjugation operation, we give a new proof for the known 
duality relation of corresponding super Schur polynomials.

In this article we also explore the question whether 
nonintegrable quantum spin chains can also exhibit duality relation 
under the exchange of bosonic and fermionic spin degrees of freedom. By 
using a mapping between the anyon like representations of the graded exchange 
operators like $\hat{P}_{jk}^{(m|n)}$ and $\hat{P}_{jk}^{(n|m)}$, we are able 
to show that the partition function of any quantum spin chain with a global 
$SU(m|n)$ symmetry would satisfy such a duality relation (\ref{e16}). This 
duality relation implies that the spectrum of the $SU(m|n)$ model can be
obtained by inverting the spectrum of the $SU(n|m)$ model and giving an overall
shift to this inverted spectrum. Since a global $SU(m|n)$ symmetry can be 
found in a wide class of integrable as well as nonintegrable spin chains, 
this duality relation seems to be a useful probe to study their spectra. 

\newpage

\section*{\begin{large} Appendix A. \end{large} 
\begin{normalsize}
Derivation of the recursion relation (\ref{c26})\end{normalsize}}
\renewcommand{\theequation}{A-\arabic{equation}}
\setcounter{equation}{0}

Let us rewrite $\tilde{S}_{\l m_1, m_2, \dots,m_r \r}(x,y)$ in 
\mbox{Eqn. (\ref{c25})} as
\bea && \tilde{S}_{\l m_1, m_2, \dots,m_r \r }(x,y) \nn \\
&& ~~=(-1)^{r-1} E_{M_r}(x,y) + \sum_{k=2}^r 
\sum_{ \{ \ell_1, \ell_2, \dots,\ell_k \}\in \mathcal{P}_{r}(k) }
(-1)^{r-k} \prod_{i=1}^k E_{M_{L_i}-M_{L_{i-1}}}(x,y) \, .~~~~ \label{A1} \eea
Since, $\{ \ell_1,\ell_2,\dots,\ell_k\} \in \mathcal{P}_r (k)$ and all these 
$\ell_s$'s are positive integers, it is evident that $\ell_k=r-\left( \ell_1+
\ell_2+\dots+\ell_{k-1}\right) \leq r-k+1$. Therefore, \mbox{Eqn. (\ref{A1})}
can also be expressed as 
\bea && \tilde{S}_{\l m_1, m_2,\dots,m_r \r }(x,y) \nn \\
&& ~=(-1)^{r-1} E_{M_r}(x,y) + \sum_{k=2}^r \sum_{\ell_k=1}^{r-k+1}
\sum_{ \{ \ell_1,\ell_2,\dots,\ell_{k-1} \} \in \mathcal{P}_{r-\ell_k}(k-1)} 
(-1)^{r-k} \prod_{i=1}^k E_{M_{L_i}-M_{L_{i-1}}}(x,y). \nn \eea
By interchanging the order of the summation variables $k$ and $\ell_k$ 
in the above equation, we obtain 
\bea && \tilde{S}_{\l m_1,m_2,\dots,m_r \r}(x,y)~~~~~~~~~~~~~~~~~~~~ \nn \\
&& =(-1)^{r-1} E_{M_r}(x,y) + \sum_{\ell_k=1}^{r-1} \sum_{k=2}^{r-\ell_k+1}
\sum_{ \{ \ell_1,\ell_2,\dots,\ell_{k-1} \} \in \mathcal{P}_{r-\ell_k}(k-1)} 
(-1)^{r-k} \prod_{i=1}^k E_{M_{L_i}-M_{L_{i-1}}}(x,y) .\nn \eea
Now, using the fact that, $L_k=\sum_{s=1}^k \ell_s =r$ and $L_{k-1}=
\sum_{s=1}^{k-1} \ell_s = r-\ell_k$, we can rewrite the above equation as
\bea && \tilde{S}_{\l m_1,m_2, \dots,m_r \r }(x,y)~~~~~~~~~~~~~~~~~~~\nn \\
&& ~~=(-1)^{r-1} E_{M_r}(x,y)+\Bigl[ \, \sum_{\ell_k=1}^{r-1}(-1)^{\ell_k-1}
E_{M_{r}-M_{r-\ell_k}}(x,y) \nn \\
&& ~~~~~~\times \sum_{k=2}^{r-\ell_k+1} \sum_{ \{ \ell_1,\ell_2,\dots,
\ell_{k-1} \} 
\in \mathcal{P}_{r-\ell_k}(k-1)}(-1)^{\ell_1+\ell_2+\dots+\ell_{k-1}-(k-1)} 
\prod_{i=1}^{k-1}E_{M_{L_i}-M_{L_{i-1}}}(x,y) \Bigr].\nn \\ \eea
Redefining $\ell_k \equiv s$ and $k-1 \equiv t$, we obtain
\bea && \tilde{S}_{\l m_1,m_2,\dots,m_r \r}(x,y) ~~~~~~~~~~~~~~~~~~~~~~~~~~~~~
~~~~~~~~~~~~~~~~~~~~~~~~~~~~~~~~~~~~~~~~~~~~~~~~~~~~~~~~ \nn \\
&& ~~=(-1)^{r-1} E_{M_r}(x,y)+\Bigl[ \, \sum_{s=1}^{r-1}(-1)^{s-1} 
E_{M_{r}-M_{r-s}}(x,y) \nn \\
&& ~~~~~~~~~~~~~ \times \sum_{t=1}^{r-s} \sum_{ \{ \ell_1,\ell_2,\dots,
\ell_{t} \} \in \mathcal{P}_{r-s}(t)} (-1)^{\ell_1+\ell_2+\dots+\ell_{t}-t}
\prod_{i=1}^{t}E_{M_{L_i}-M_{L_{i-1}}}(x,y) \Bigr]. ~~~~~~~~~~~~~ \eea
Finally, by using the above equation and the definition of 
$\tilde{S}_{\l m_1,m_2,\dots,m_{r-s} \r}(x,y)$ from \mbox{Eqn. (\ref{c25})}, we
derive the recursion relation for $\tilde{S}_{\l m_1,m_2,\dots,m_r \r}(x,y)$ as
\bea \tilde{S}_{ \l m_1,m_2,\dots,m_r \r }(x,y) \hskip -.2cm &=& \hskip -.22cm 
(-1)^{r+1} E_{M_r}(x,y)+ \! \sum_{s=1}^{r-1}(-1)^{s+1} E_{M_{r}-M_{r-s}}(x,y)
\cdot \tilde{S}_{ \l m_1,m_2,\dots,m_{r-s} \r }(x,y) \nn \\
\hskip -.2cm &=& \hskip -.22cm
\sum_{s=1}^{r} (-1)^{s+1} E_{m_r+m_{r-1}+\dots+m_{r-s+1}}(x,y)
\cdot \tilde{S}_{ \l m_1,m_2,\dots,m_{r-s} \r }(x,y), \eea
where $\tilde{S}_{\l 0 \r }(x,y)=1$.

\end{document}